\documentclass[final,3p,times,11pt]{elsarticle}

\usepackage{amsmath, amssymb}
\usepackage{lipsum}
\journal{Nuclear Physics B}
\usepackage{graphicx}
\usepackage[caption=false]{subfig}
\usepackage{dcolumn}
\usepackage{bm}
\usepackage{natbib}
\usepackage{url}

\usepackage{tikz}
\usetikzlibrary{shapes}
\usetikzlibrary{positioning}
\usetikzlibrary{calc} 
\usepackage{qcircuit}
\usepackage{braket}

\usepackage{ulem}
\usepackage[colorlinks=true,backref=false, linktocpage=true,
citecolor=blue,urlcolor=blue,linkcolor=blue,pdfpagemode=UseOutlines]{hyperref}
\usepackage{todonotes}
\usepackage{multirow}
\usepackage[section]{placeins}
\usepackage{booktabs}
\usepackage{appendix}

\begin{document}

\begin{frontmatter}

\title{Capturing the Page Curve and Entanglement Dynamics of Black Holes in Quantum Computers}

\author[first,second]{Talal Ahmed Chowdhury}
\address[first]{Department of Physics, University of Dhaka, P.O. Box 1000, Dhaka, Bangladesh.}
\address[second]{Department of Physics and Astronomy, University of Kansas, Lawrence, Kansas 66045, USA.}
\ead{talal@ku.edu}
\author[third]{Kwangmin Yu}
\address[third]{Computational Science Initiative, Brookhaven National Laboratory, Upton, New York 11973, USA.}
\ead{kyu@bnl.gov}
\author[fourth]{Muhammad Asaduzzaman}
\address[fourth]{Department of Physics and Astronomy, University of Iowa, Iowa City, Iowa 52242, USA.}
\author[fifth,sixth,seventh]{Raza Sabbir Sufian}
\address[fifth]{Department of Physics, New Mexico State University, Las Cruces, NM 88003, USA}
\address[sixth]{RIKEN-BNL Research Center, Brookhaven National Laboratory, Upton, New York 11973, USA.}
\address[seventh]{Physics Department, Brookhaven National Laboratory, Upton, New York 11973, USA.}
\ead{gluon2025@gmail.com}

\begin{abstract}
Quantum computers are emerging technologies expected to become important tools for exploring various aspects of fundamental physics in the future. Therefore, we pose the question of whether quantum computers can help us to study the Page curve and the black hole information dynamics, which has been a key focus in fundamental physics. In this regard, we rigorously examine the qubit transport model, a toy qubit model of black hole evaporation on IBM's superconducting quantum computers, to shed light on this question. Specifically, we implement the quantum simulation of the scrambling dynamics in black holes using an efficient random unitary circuit. Furthermore, we employ the swap-based many-body interference protocol and the randomized measurement protocol to measure the entanglement entropy of Hawking radiation qubits in this model. Finally, by incorporating quantum error mitigation techniques into our challenging implementation of entanglement entropy measurement protocols on the IBM quantum hardware, we accurately determine the R\'enyi entropy in the qubit transport model, thus showcasing the utility of quantum computers for future investigations of complex quantum systems.
\end{abstract}

\end{frontmatter}

\section{Introduction}\label{sec:intro}
Quantum entanglement \cite{horodecki} is paramount in quantum computing, quantum information science, quantum many-body physics, high energy and nuclear physics, and the physics of black holes. Its dynamics are tied to the non-equilibrium aspects and thermalization of quantum many-body systems \cite{Eisert-1, Gogolin-1}. Besides, harnessing the dynamics of quantum information and entanglement is essential for developing quantum technologies and studying complex quantum systems \cite{Jozsa-entanglement, Preskill-entanglement}.

Recently, significant experimental progress has been made to measure entanglement entropy and characterize quantum systems \cite{Islam-entanglement-swap, Kaufman-entanglement-swap, Brydges-randomized, Lukin-randomized-MBL, Cheneau-light-cone-spreading, Abanin-Quantum-switch, Daley-ent-growth-quench, Richerme-entanglement-propagation, Jurecevic-entanglement-propagation, Schweigler-higher-order-corr} (for a review, see Ref.~\cite{Swan-dynamics-quantum-info-review}). Quantum computers, with their impressive advancement over the past years, have positioned themselves as an important experimental tool offering in-depth understanding of the quantum many-body systems, as the fundamental building blocks of the many-body systems either have the same degrees of freedom or can be easily mapped to the qubits/qudits of the quantum computers. Moreover, the quantum computer as the quantum simulator \cite{cirac-zollar, Georgescu:2013oza, Daley:2022eja} can probe not only the microscopic intricacies of the quantum systems but also the emergent collective phenomena not seen in their constituents. Besides, due to the exponential scaling of the Hilbert space when the system size increases, the entanglement entropy computation in classical computers becomes intractable for general highly entangled states. Therefore, quantum computers can be a perfect means to measure the entanglement entropy of quantum many-body systems. 

Quantum entanglement also plays a crucial role in the Hawking radiation and the black hole information \cite{Hawking:1975vcx, Page:1979tc, Page-BH-information, Page-information-in-BH-radiation, Giddings:2006sj, Page:2004xp, Mathur:2008wi, Page:2013dx, Entropy-of-Hawking-radiation}. Inspired by the Hayden-Preskill thought experiment \cite{Hayden-Preskill} (also known as Hayden-Preskill protocol) and Yoshida-Kitaev decoding protocol \cite{Yoshida-Kitaev}, both addressing the information retrieval and reconstruction of the quantum state from the Hawking radiation, we set to present how a quantum computer can probe the entanglement dynamics of the evaporating black hole. For this purpose, we consider a toy qubit model of the black hole, namely the qubit transport model \cite{Osuga:2016htn}  which describes the unitary dynamics of the black hole evaporation and information transfer to the Hawking radiation (similar models are \cite{Giddings:2011ks, Giddings:2012bm, Giddings:2012dh, Nomura:2014woa}). In our quantum simulation of the qubit transport model on IBM's superconducting quantum computers, the entropy of the Hawking radiation entails gradually measuring the entanglement entropy of qubit subsystems of different sizes. 

We know that when the quantum state is in the random pure state, the averaged entanglement entropy of its subsystems, taken with respect to unitary-equivalent Haar measure, is given by the Page entropy \cite{Page:1993df, Lubkin:1978nch, Lloyd:1988cn, Sen:1996ph}. The Page entropy captures the universal feature of random pure states relevant to the unitary dynamics of the black holes and quantum many-body systems. In this work, we present a way to utilize the quantum computer to measure the Page entropy of subsystems of such a Haar-random quantum state. We use two protocols: the swap-based many-body interference (SWAP-MBI) protocol \cite{Buhrman:2001rma, Horodecki-swap, Ekert-swap} and the randomized measurement (RM) protocol \cite{Brydges-randomized, Enk-Beenakker, Elben-1, Vermersch-1, Elben-2, Rath, Elben-3}, to measure the R\'enyi entropy of the subsystems, which we called the Hawking radiation qubits in later sections. We consider the Haar-random quantum state to represent the highly entangled state of the black hole qubits in our black hole quantum simulation setup. We mainly focus on the build-up of R\'enyi entropy in those subsystems under the random unitary dynamics, thus simulating the scrambling of information in black holes. Such random unitary dynamics in our quantum simulation are carried out using an efficient random unitary circuit (RUC) of the brickwall structure (for a review, see Ref.~\cite{random-unitary-circuits}) with quantum computers with limited qubit connectivity in mind. Furthermore, we minimize the circuit depth of each brickwall layer by adopting an optimal two-qubit entangling gate \cite{zhang2024optimal}. We implement our protocols on IBM's superconducting quantum computers. Using quantum error mitigation techniques, we measure the entanglement entropy of various subsystems. This allows us to leverage quantum computers to directly explore the Page curve. The Page curve is closely related to black hole dynamics and quantum information scrambling in quantum many-body systems.

The structure of the manuscript is as follows. In section \ref{sec:qubit-trasport-main}, we describe the Page entropy of a quantum system, the qubit transport model for the unitary black hole evaporation, and its quantum circuit implementation. Section \ref{sec:scrambling} describes the implementation of black hole's scrambling dynamics using our efficient random unitary circuit. In Section \ref{sec:QC-EE}, we outline the implementation of two protocols for measuring entanglement (R\'enyi) entropy on quantum computers: the swap-based many-body interference protocol and the randomized measurement protocol, designed with IBM's superconducting quantum computers in mind. We present the results from our numerical simulation and experiments on actual quantum computer in section \ref{sec:result}. Finally, we conclude in section \ref{sec:conclusion}.

\section{Qubit transport model of the Black hole in Quantum Computers}\label{sec:qubit-trasport-main}

\subsection{Black hole as a quantum system}\label{sec:black hole as quantum}
\subsubsection{Page entropy}\label{sec:page entropy}
First, we introduce the general notion of entanglement entropy as it is associated with a quantum system. Consider a system of $N$ qubits in a random pure state $|\psi\rangle$, whose density matrix is given by $\rho=|\psi\rangle\langle\psi|$. If we partition this system into two subsystems of $L$ and $M=N-L$ qubits, the entanglement entropy, i.e., the $n$-R\'enyi entropy associated with the $L$-qubit subsystem is given by,
\begin{equation}
    S^{(n)}_{L} = \frac{1}{1-n}\mathrm{log}[\mathrm{Tr}(\rho_{L}^{n})]
    \label{eq:n-renyi}
\end{equation}
where $\rho_{L}$ is the reduced density matrix of the subsystem with $L$ qubits, obtained by tracing the subsystem of $M = N-L$ qubits, $\rho_{L}=\mathrm{Tr}_{M}\rho$. The $n = 2$ case is the 2-R\'enyi entropy (R\'enyi entropy for simplicity), whose measurement we are focusing on in this study. Additionally, when $n\rightarrow 1$, the $n$-R\'enyi entropy boils down to the von Neumann entropy,
\begin{equation}
    S^{vN}_{L} = -\mathrm{Tr}\rho_{L}\mathrm{log}\rho_{L}.
    \label{eq:von-neumann}
\end{equation}

Now the Page entropy, i.e. the average von Neumann entropy for the $L$-qubit subsystem over all random pure states $\rho$ is given by
\begin{equation}
    S^{vN}_{L,M} = \sum_{k = 2^{M}+1}^{2^N}\frac{1}{k}-\frac{2^L-1}{2^{M+1}},\,\,\,\, L\leq M
    \label{eq:page-entropy}
\end{equation}
where, the average is taken with respect to unitarily invariant Haar measure on the space of unit vectors $|\psi\rangle$ in the $2^{N}$ dimensional Hilbert space \cite{Page:1993df, Sen:1996ph}. Besides, the space of such random states is equivalent to the volume of hypersphere $S^{2^{N+1}-1}$. For $L>M$ case, we make the following substitution $(L\leftrightarrow M)$ in Eq. (\ref{eq:page-entropy}). When $1\ll L$, it simplifies to
\begin{equation}
    S^{vN}_{L,M}\approx L\,\mathrm{log}2 - (1/2^{N-2L-1})
    \label{eq:page-simple}
\end{equation}
which further becomes the thermodynamic entropy for the $L$-qubit subsystem, $S^{\mathrm{Th}}_{L}\approx L\,\mathrm{log}2$ when $L\ll N$.

On the other hand, the average R\'enyi entropy corresponding to the $L$-qubit subsystem over the space of random states $|\psi\rangle$ is given by \cite{Liu:2017asn},
\begin{equation}
    S^{(2)}_{L} = -\mathrm{log}\left(\frac{2^{L}+2^{N-L}}{2^{N}+1}\right).
    \label{eq:renyi-entropy}
\end{equation}
Moreover, as the $n$-R\'enyi entropy satisfies the inequality\footnote{For a detailed discussion, see Ref. \cite{Liu:2017lem}} $S^{(m)}_{L}\geq S^{(n)}_{L}$ for $m\leq n$ for a $L$-qubit subsystem, we can see that the Page entropy, namely the von Neumann entropy given in Eq. (\ref{eq:page-entropy}) will serve as the upper bound of the corresponding R\'enyi entropy $S^{(2)}_{L}$ of the corresponding subsystem.

\subsubsection{The entropy of the Hawking radiation}\label{sec:Hawking radiation entropy}
We now provide a brief overview of the entropy of Hawking radiation as proposed by Page~\cite{Page-information-in-BH-radiation}. The Bekenstein-Hawking (BH) entropy \cite{Bekenstein:1973ur, Hawking:1971tu}, $S_{\mathrm{BH}}$ associated with a black hole in Planck units ($\hbar = c = G=k_{\mathrm{Boltzmann}}=1$) is given by,
\begin{equation}
    S_{\mathrm{BH}}=\frac{A}{4} .
    \label{eq:BH entropy}
\end{equation}
where, $A$ is the area of the black hole. Due to the finiteness of $S_{\mathrm{BH}}$, we consider, assuming the entropy to be an integer, that the black hole consists of $N=S_{\mathrm{BH}}/\mathrm{ln}2$ qubits. Moreover, we assume that the quantum state associated with $N$ qubits is in a pure state after the formation of the black hole. However, the qubits themselves are highly entangled with each other due to the fast scrambling governed by a complex unitary transformation.
\begin{figure}
    \centering
    \includegraphics[width=0.5\textwidth]{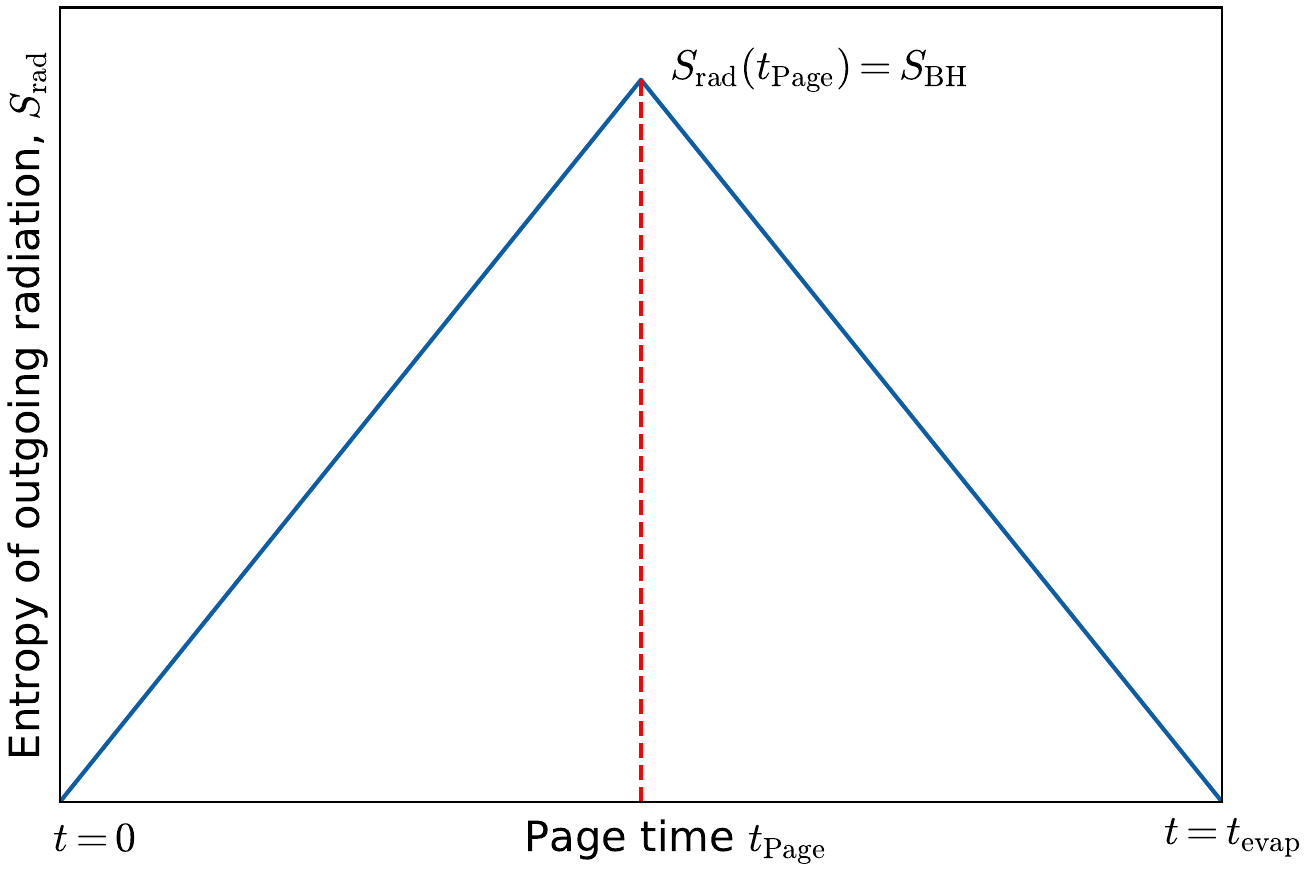}
    \caption{Schematic behavior of the entropy of the outgoing Hawking radiation when the black hole is governed by unitary dynamics. This curve is known as the Page curve and is associated with the entropy of the Hawking radiation. }
    \label{fig:page-curve}
\end{figure}
Similarly, the outgoing Hawking radiation can be thought of as made out of qubits. As the black hole evaporates gradually by the emission of Hawking radiation, in our qubit scenario of the black hole, this physical process is represented by the reduction of the black hole qubits $N$ and the increase of Hawking radiation qubits denoted as $N_{\mathrm{rad}}$. During the emission of the Hawking radiation, the Hawking radiation qubits will have a varying entanglement with the rest of the black hole qubits. At a specific time called the Page time, $t_{\mathrm{Page}}$, the entanglement entropy $S_{\mathrm{rad}}$ of the outgoing Hawking qubits will be equal to the Bekenstein-Hawking entropy of the remaining black hole. At $t_{\mathrm{Page}}$, the Hawking qubits will be maximally entangled with the remaining black hole qubits. After the Page time, the entanglement entropy of the Hawking qubits will gradually decrease because the information content of the black hole beyond the horizon is now transferred more and more to the Hawking qubits. When the black hole evaporates, all of its information content is transferred to the Hawking qubits, which are again in a pure state and thus will have zero entanglement entropy. In Fig. \ref{fig:page-curve}, we present the above schematic picture of the time dependence of the entanglement entropy associated with the Hawking radiation. In the subsequent sections, we will elaborate on these dynamics with a more microscopic but still toy model of a black hole with qubits, namely the qubit transport model of the black hole.

\subsection{Qubit Transport Model}\label{sec:qubit transport model}
In the qubit transport model of the black hole, we assume the black hole consists of $N$ qubits, which are entangled due to the fast scrambling. Apart from this assumption, we consider $N$ outgoing radiation qubits just outside and $N$ infalling radiation qubits just inside the black hole's event horizon. Furthermore, an infalling radiation qubit and an outgoing radiation qubit form a singlet Bell state, and the vacuum state at the vicinity of the event horizon can be considered as the tensor product of such a singlet Bell state. As the $2 N$ infalling and outgoing radiation qubits at the horizon are part of the vacuum fluctuations, as in the case of forming the Hawking radiation, the product of the singlet Bell state does not contribute to the entropy of the black hole. We denote the black hole qubits, infalling radiation qubits, and outgoing radiation qubits as $q_{i}$, $a_{i}$ and $b_{i}$, respectively. The singlet Bell state of the infalling and outgoing radiation qubits is,
\begin{equation}
    |B\rangle_{a_{i}b_{i}}=\frac{1}{\sqrt{2}}(|0\rangle_{a_{i}}|1\rangle_{b_{i}}-|1\rangle_{a_{i}}|0\rangle_{b_{i}}) .
    \label{eq:Bell}
\end{equation}

Initially the quantum state of the black hole and radiation qubits is given by
\begin{align}
    |\Psi_{\mathrm{initial}}\rangle =& \left(\sum_{s_{1},s_{2},..,s_{N}=0}^{1}A_{s_{1}s_{2}..s_{N}}|s_{1}\rangle_{q_{1}}|s_{2}\rangle_{q_{2}}...|s_{N}\rangle_{q_{N}}\right)
    \otimes\prod_{i=1}^{N}|B\rangle_{a_{i}b_{i}} 
    \label{eq:BH state}
\end{align}
where $|s_{i}\rangle_{q_{i}}$ is the quantum state of the $i$-th black hole qubit $q_i$ in the computational basis.

In the qubit transport model, it is assumed that the quantum state of $i$-th black hole qubit $q_{i}$ and the singlet Bell state associated with the $i$-th infalling and outgoing radiation qubits undergo the following transition,
\begin{equation}
    |s_{i}\rangle_{q_{i}}|B\rangle_{a_{i}b_{i}}\rightarrow -|B\rangle_{q_{i}a_{i}}|s_{i}\rangle_{b_{i}}
    \label{eq:swap operation}
\end{equation}
which can be realized by the continuous unitary transformation $U(\theta)$ on the state $|s_{i}\rangle_{q_{i}}|B\rangle_{a_{i}b_{i}}$ where,
\begin{align}
    U(\theta) &= e^{-i\theta P_{q_{i}a_{i}b_{i}}}\,\,\,\mathrm{with}\nonumber\\
    P_{q_{i}a_{i}b_{i}}&=\frac{1}{2}\left[|0\rangle_{q_{i}}\langle 0|_{q_{i}}I_{a_{i}}|1\rangle_{b_{i}}\langle 1|_{b_{i}}-|1\rangle_{q_{i}}\langle 0|_{q_{i}}I_{a_{i}}|0\rangle_{b_{i}}\langle 1|_{b_{i}}
    -|0\rangle_{q_{i}}\langle 1|_{q_{i}}I_{a_{i}}|1\rangle_{b_{i}}\langle 0|_{b_{i}}+|1\rangle_{q_{i}}\langle 1|_{q_{i}}I_{a_{i}}|0\rangle_{b_{i}}\langle 0|_{b_{i}}\right]
    \label{eq:theta-swap}
\end{align}
At $\theta = \pi$ the unitary transformation $U(\pi) = \mathbb{I}-2P_{q_{i}a_{i}b_{i}}$ executes the swap operation as in Eq. (\ref{eq:swap operation}). One can consider the $\theta$ parameter of $U(\theta)$ is a function of radius $r$ that changes 0 at the black hole horizon to $\pi$ at the radial infinity. It can be also expressed as
\begin{equation}
    \theta = \pi\left(1-K/K_{\mathrm{horizon}}\right)
    \label{eq:theta parameter}
\end{equation}
where $K=R^{\mu\nu\rho\sigma}R_{\mu\nu\rho\sigma}$ is the Kretschmann scalar \cite{Cherubini:2002gen} associated with the Riemann curvature tensor $R^{\mu}_{\nu\rho\sigma}$ as it captures the curvature of the space-time with radial distance from the black hole horizon, and $K_{\mathrm{horizon}}$ is its value at the horizon. Therefore, we can see that due to the swap operation between the quantum states of the black hole qubit $q_{i}$ and the outgoing radiation qubit $b_{i}$, the black hole qubit's quantum state emits through the Hawking radiation qubit. After the emission, the remaining singlet Bell state of the black hole qubit $q_{i}$ and infalling radiation qubit $a_{i}$ can be treated as the part of the vacuum state, so effectively the black hole will now have $N-1$ black hole qubits. One can notice that although the Hawking radiation qubit is emitted from the black hole, its quantum state has been entangled with the rest of the black hole qubits due to the fast scrambling. For this reason, the Hawking radiation qubits will maintain the entanglement with the black hole qubits. As a consequence, after the emission of the $N_{\mathrm{rad}}$ Hawking radiation qubits, the quantum state of the effective $N-N_{\mathrm{rad}}$ black hole qubits and $N_{\mathrm{rad}}$ Hawking radiation qubits will still be entangled. If we trace out the subsystem of the $N-N_{\mathrm{rad}}$ black hole qubits, as one would never have any direct access to them, we can immediately compute the entanglement entropy of $N_{\mathrm{rad}}$ Hawking qubits given by the quantity $S_{\mathrm{rad}}$.

Finally, when all of $N$ of the original outgoing radiation qubits have left the black hole and propagated to become the Hawking radiation qubits, there are no qubits left for the black hole; hence, it is completely evaporated away. The quantum state associated with the initial $N$ entangled black hole qubits is now transferred into $N$ Hawking radiation qubits, which then form a pure quantum state, just like the original quantum state of the black hole qubits, as follows,
\begin{equation}
    |\Psi_{\mathrm{final}}\rangle = \sum_{s_{1},s_{2},..,s_{N}=0}^{1}A_{s_{1}s_{2}...s_{N}}|s_{1}\rangle_{b_{1}}|s_{2}\rangle_{b_{2}}...|s_{N}\rangle_{b_{N}} .
    \label{eq:final Hawking state}
\end{equation}

\subsection{Quantum Circuit for the Qubit Transport Model}\label{sec:QC for qubit transport}
\begin{figure}[t!]
\centerline{\includegraphics[width=18cm]{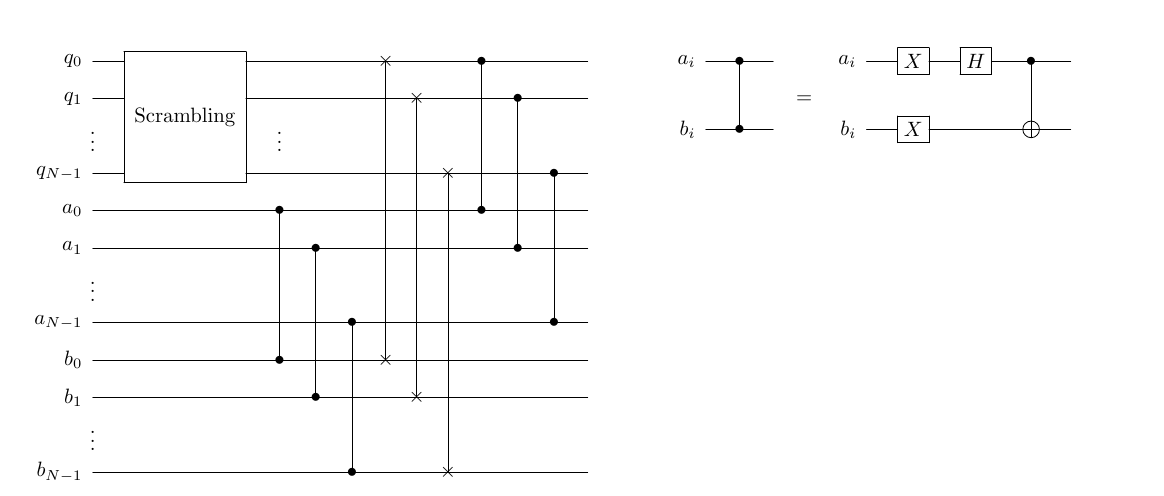}
}
\caption{The $N$ black hole qubits, infalling qubits, and outgoing qubits are denoted by $q_{i}$, $a_{i}$ and $b_{i}$, respectively, in the left quantum circuit. The $i$-th infalling and outgoing qubits form the singlet Bell state $|B\rangle_{a_{i}b_{i}}$, represented in the quantum circuit at right. Now, the outgoing qubits $b_{i}$ interact with the highly entangled black hole qubits $q_{i}$ through the swap gate. As a result, the quantum states of the black hole qubits $q_{i}$ are now transferred to the outgoing Hawking radiation qubits. The resulting Bell states $|B\rangle_{q_{i}a_{i}}$ become part of the vacuum fluctuations and can be omitted from further computations.}
\label{fig:qubit-transport}
\end{figure}

\begin{figure}[h!]
\centerline{\includegraphics[width=18cm]{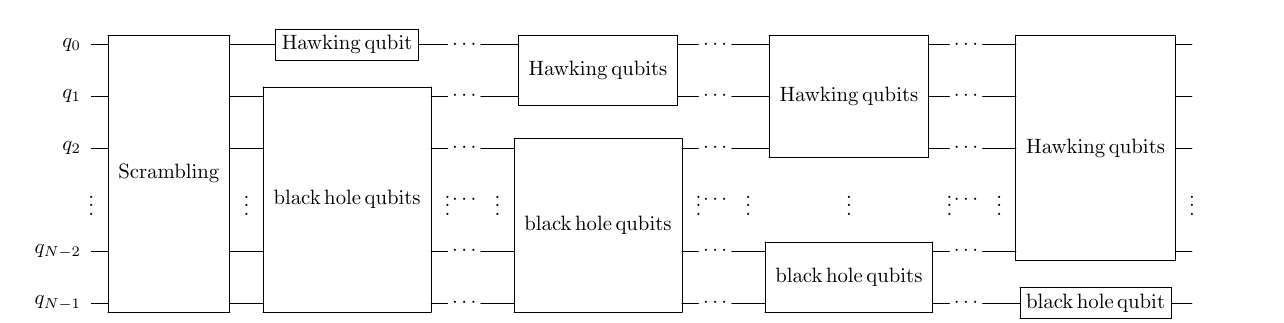}
}
\caption{Another viewpoint of the qubit transport model in the quantum computer. Here, the scrambling operation prepares a highly entangled state of $N$ black hole qubits. Then, the subsystems consist of $q_{0}$, $\{q_{0},q_{1}\}$, $\{q_0, q_{1}, q_{2}\}$ qubits, and so on,  are considered as the Hawking radiation qubits.}
\label{fig:compact qubit transport}
\end{figure}

One can represent the previous discussion of the qubit transport model in terms of quantum circuits. The initial representation of the model with the quantum circuit requires three $N$-qubit registers realizing the black hole qubits $q_{i}$, infalling radiation qubits $a_{i}$ and outgoing radiation qubits $b_{i}$, respectively, as seen in Fig. \ref{fig:qubit-transport}. We can implement the scrambling dynamics of the black hole dynamics with the random unitary circuit or time evolution operator driven by some Hamiltonian. The explicit implementation of this scrambling dynamics is addressed in section \ref{sec:scrambling}. In addition, the vacuum state is the product of $N$ singlet Bell states $|B\rangle_{a_{i}b_{i}}$, each formed through the pairing of an infalling and an outgoing radiation qubit, $a_{i}$ and $ b_{i}$, respectively as shown in Fig. \ref{fig:qubit-transport} (right). At this point, the quantum state associated with our quantum circuit represents the total quantum state of the black hole and its vacuum fluctuation $|\Psi_{\mathrm{initial}}\rangle$ in Eq. (\ref{eq:BH state}). Now, instead of considering the unitary transformation given in Eq. (\ref{eq:theta-swap}) with the $\theta$ parameter depending on the curvature with respect to the radial distance from the horizon, we directly implement Eq. (\ref{eq:swap operation}) through the consecutive swap gates between $i$-th black hole qubit $q_{i}$ and the outgoing radiation qubit $b_{i}$. This operation will lead to the transfer of the black hole quantum state in the black hole qubit register $\{q_{0},...,q_{N-1}\}$ to the outgoing radiation qubit register $\{b_{0},...,b_{N-1}\}$ which is now the Hawking radiation qubits and the product state of the singlet Bell states $|B\rangle_{q_{i}a_{i}}$ between $i$-th black hole qubit $q_{i}$ and the infalling radiation qubit $a_{i}$ can be considered as the part of vacuum fluctuations. As such product state of the singlet Bell states is not used in overall computation, we can realize the physical aspects of the qubit transport model in a more compact way using the quantum circuit in Fig. \ref{fig:compact qubit transport}. 

In this case, we again implement the scrambling on the black hole qubits $q_{i}$. Now we select the subsystems of $N_{\mathrm{rad}}=1,2,.., N$ qubits $\{q_{0},...,q_{N_{\mathrm{rad}}}\}$ as the Hawking radiation qubits in a subsequent manner as shown in Fig. \ref{fig:compact qubit transport}. Due to the scrambling, all black hole qubits are entangled. Therefore, for every subsystem of $N_{\mathrm{rad}}$ qubits with increasing $N_{\mathrm{rad}}$, the corresponding reduced density matrix will give rise to the entanglement entropy, i.e., the R\'enyi entropy $S^{(2)}$, of the Hawking radiation qubits. The gradual increase of the subsystem size of Hawking radiation qubits $N_{\mathrm{rad}}$ can be identified with the time of the black hole evolution as shown in Fig. \ref{fig:page-curve}. Therefore, in our simplified quantum circuit dynamics, the Page time $t_{\mathrm{Page}}$ will imply the maximum value of the entanglement entropy $S^{\mathrm{max}}_{\mathrm{rad}}=S_{\mathrm{BH}}$ for the subsystem of $N_{\mathrm{rad}}=N/2$ Hawking qubits. For $t>t_{\mathrm{Page}}$, the subsystem of Hawking qubits will have $N_{\mathrm{rad}}>N/2$, which leads to the decrease of its corresponding entanglement entropy because the information content of the Hawking qubit subsystem increases. Eventually, at $t=t_{\mathrm{evap}}$, the Hawking radiation subsystem will contain $N_{\mathrm{rad}}=N$ qubits and, therefore, will represent again the pure state of the black hole qubits.

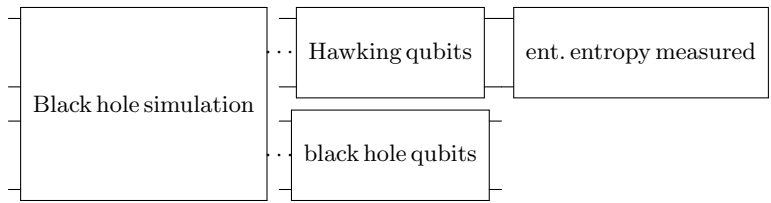
\begin{figure}[h!]
\centerline{
\Qcircuit @C=0.5em @R=0.5em @!R{
\lstick{}&\multigate{5}{\mathrm{Black\,hole\, simulation}}& & \multigate{2}{\mathrm{Hawking\, qubits}} & \qw &\multigate{2}{\mathrm{ent.\,entropy\,measured}}\\
 & &\cdots & & & \\
\lstick{}&\ghost{\mathrm{Black\,hole\, simulation}} & &\ghost{\mathrm{Hawking\,qubits}} &\qw &\ghost{\mathrm{ent.\,entropy\, measured}}\\
\lstick{}&\ghost{\mathrm{Black\,hole\,simulation}} & &\multigate{2}{\mathrm{black\,hole\, qubits}} & \qw & &\\
& &\cdots & & & &\\
\lstick{}&\ghost{\mathrm{Black\,hole\,simulation}} & &\ghost{\mathrm{black\,hole\,qubits}} & \qw & &
}
}
\caption{One can also imagine simulating the black hole dynamics, namely its scrambling dynamics in the quantum computer and measure the entanglement entropy of the subsystems using the protocols.}
\label{fig:QC-for-BH-2}
\end{figure}

As mentioned in section \ref{sec:intro}, quantum computers are the perfect means of measuring the entanglement entropy of a subsystem. Therefore, we can adopt two possible uses of quantum computers in studying the entanglement entropy of the black hole.
\begin{itemize}
    \item An observer near the black hole continuously collects the Hawking radiation qubits right after its formation. The access to a quantum computer immediately allows the observer to measure the entanglement entropy associated with collected Hawking radiation qubits. Even if there is no accessible quantum computer nearby, the quantum state can be transmitted to an available one using the teleportation protocol. Thus, the quantum computer can be utilized to study the entanglement entropy of the Hawking radiation qubits directly. We present a schematic diagram of this viewpoint in Fig. \ref{fig:QC-for-BH-1}. Furtherore, due to noise, the fidelity of teleportation will deviate from $ F = 1 $. However, the entanglement entropy associated with the teleported state depends on the nature of the noise. For example, in the case of depolarizing noise, the imperfect teleportation with $ F < 1 $ will shift the teleported state towards a more maximally mixed state. In contrast, the state of the $ L$ Hawking radiation qubits will generally be maximally mixed when we trace out the $ N - L $ black hole qubits from the pure state of $ N $ black hole qubits, which are in a highly entangled state formed by black hole scrambling (essentially a Haar-random unitary operation). Therefore, imperfect teleportation caused by depolarizing noise will have small affect in the entanglement entropy of the teleported state. On the other hand, if we encounter noise such as amplitude damping, the teleported state will be pushed towards one with greater purity, resulting in a smaller measured entanglement entropy.

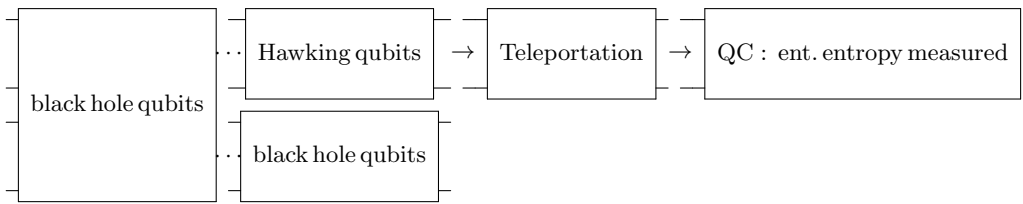
\begin{figure}[h!]
\centerline{
\Qcircuit @C=0.5em @R=0.5em @!R{
\lstick{}&\multigate{5}{\mathrm{black\,hole\,qubits}}& & \multigate{2}{\mathrm{Hawking\, qubits}} & \qw & &\qw &\multigate{2}{\mathrm{Teleportation}}& \qw & &\qw &\multigate{2}{\mathrm{QC:\,ent.\,entropy\,measured}}\\
& &\cdots & & &\rightarrow & & & &\rightarrow & &\\
\lstick{}&\ghost{\mathrm{black\,hole\,qubits}} & &\ghost{\mathrm{Hawking\,qubits}}& \qw & &\qw &\ghost{\mathrm{Teleportation}} &\qw & &\qw &\ghost{\mathrm{QC:\,ent.\,entropy\, measured}}\\
\lstick{}&\ghost{\mathrm{black\,hole\,qubits}} & &\multigate{2}{\mathrm{black\,hole\, qubits}} & \qw & & & & & & & &\\
& &\cdots & & & & & & & & & &\\
\lstick{}&\ghost{\mathrm{black\,hole\,qubits}} & &\ghost{\mathrm{black\,hole\,qubits}} & \qw & & & & & & & &
}
}
\caption{We could imagine collecting the Hawking radiation, thereby its quantum state, and using the quantum teleportation protocol to send it to the quantum computer, which can then be used to measure its entanglement entropy using the entanglement entropy measurement protocols.}
\label{fig:QC-for-BH-1}
\end{figure}
    
    \item As the qubit transport model considers the unitary dynamics of the black hole, one can use the quantum computer to carry out the quantum simulation of the black hole's scrambling dynamics. As sketched in Fig. \ref{fig:compact qubit transport} and \ref{fig:QC-for-BH-2}, the entanglement dynamics of the Hawking radiation qubits can be captured consecutively by measuring the entanglement entropy of qubit subsystems of increasing sizes.  
\end{itemize}

\section{Scrambling dynamics of Black Holes in Quantum Computers}\label{sec:scrambling}
\begin{figure}[t!]
\[
\Qcircuit @C=1em @R=0.7em {
\lstick{ {q}_{0} :  } & \qw & \multigate{1}{SU(4)} & \qw & \qw & \qw & & \cdots & & & \multigate{1}{SU(4)} & \qw & \qw & \qw  \\
\lstick{ {q}_{1} :  } & \qw & \ghost{SU(4)} & \qw  & \multigate{1}{SU(4)} & \qw & & \cdots & & & \ghost{SU(4)} & \qw  & \multigate{1}{SU(4)} & \qw \\
\lstick{ {q}_{2} :  } & \qw & \multigate{1}{SU(4)} & \qw  & \ghost{SU(4)} & \qw & & \cdots & & & \multigate{1}{SU(4)} & \qw  & \ghost{SU(4)} & \qw \\
\lstick{ {q}_{3} :  } & \qw & \ghost{SU(4)} & \qw & \multigate{1}{SU(4)} & \qw & & \cdots&  & & \ghost{SU(4)} & \qw & \multigate{1}{SU(4)} & \qw \\
\lstick{ {q}_{4} :  } & \qw & \multigate{1}{SU(4)} & \qw & \ghost{SU(4)} & \qw & & \cdots & & & \multigate{1}{SU(4)} & \qw & \ghost{SU(4)} & \qw \\
\lstick{ {q}_{5} :  } & \qw & \ghost{SU(4)}& \qw & \qw & \qw & & \cdots & & & \ghost{SU(4)}& \qw & \qw & \qw \\
\vdots  &  & \vdots & & \vdots & & &\ddots & & & \vdots & & \vdots \\
\lstick{ {q}_{N-4} :  } & \qw & \multigate{1}{SU(4)} & \qw & \qw & \qw & & \cdots & & & \multigate{1}{SU(4)} & \qw & \qw & \qw \\
\lstick{ {q}_{N-3} :  } & \qw & \ghost{SU(4)}& \qw & \multigate{1}{SU(4)} & \qw & & \cdots & & & \ghost{SU(4)}& \qw & \multigate{1}{SU(4)} & \qw \\
\lstick{ {q}_{N-2} :  } & \qw & \multigate{1}{SU(4)} & \qw & \ghost{SU(4)} & \qw & & \cdots & & & \multigate{1}{SU(4)} & \qw & \ghost{SU(4)} & \qw \\
\lstick{ {q}_{N-1} :  } & \qw & \ghost{SU(4)}& \qw & \qw & \qw & & \cdots & & & \ghost{SU(4)}& \qw & \qw & \qw
\gategroup{1}{3}{11}{3}{.7em}{-}
\gategroup{1}{4}{11}{6}{.7em}{--}
\gategroup{1}{11}{11}{11}{.7em}{-}
\gategroup{1}{12}{11}{14}{.7em}{--}
}
\]
\caption{The brickwall circuit of Haar-random unitary operation. The layers surrounded by the straight lines are the even layers and the layers surrounded by the dotted lines are the odd layers.}
\label{fig:random-circuit}
\end{figure}
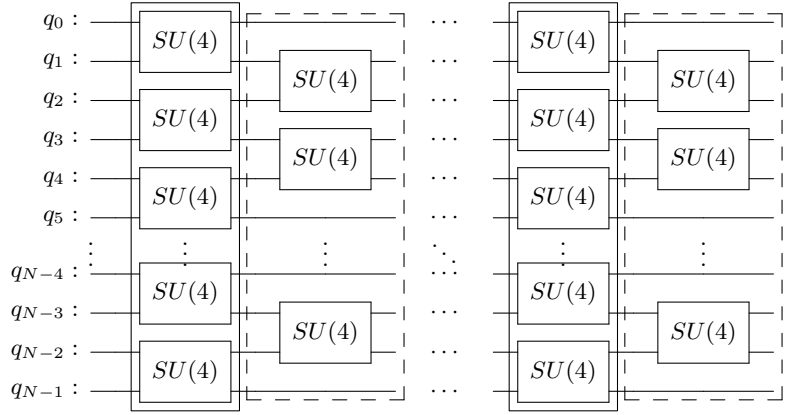

We simulate the system of $N$ black hole qubits with the one-dimensional spin chain with open boundary conditions, and the scrambling dynamics is implemented by the random unitary circuit (RUC) acting on an initial product state $|0\rangle^{\otimes N}$, as shown in Fig. \ref{fig:random-circuit}. We adopt the brickwall construction for the RUC to approximate the Haar-random unitary operator acting on $N$ qubits. In this construction, a product of two-qubit unitary gates, denoted by $U_{\mathrm{even}}$, where each two-qubit unitary gate is sampled from $SU(4)$ Haar measure, acts on the pairs of neighboring qubits with even bonds in the spin chain, followed by another product of two-qubit unitary gates acting on qubit pairs of odd bonds, and denoted by $U_{\mathrm{odd}}$.

Therefore, the Haar-random unitary operator acting on the spin chain is given by
\begin{equation}
    U_{\mathrm{RUC}} = \prod_{i=1}^{\mathrm{Layers}} (U_{\mathrm{odd}}\,U_{\mathrm{even}})_{i}
    \label{eq:brickwall}
\end{equation}
where a single layer in the brickwall RUC is given by, $U_{1-\mathrm{Layer}}=U_{\mathrm{odd}}U_{\mathrm{even}}$. Immediately we can see that, the entanglement between the $N$ qubits will grow with the increasing number of brickwall layers. In the subsequent sections, we will study the entanglement entropy of the subsystems by applying the entanglement entropy measuring protocols in the quantum computer.

Each $SU(4)$ two-qubit gate represents $4\times 4$ unitary matrix $U^{(4)}$, sampled from the Haar measure that can be decomposed using Cartan-Weyl decomposition \cite{KHANEJA200111, Zhang-Vala-two-qubit, Zhang:2003zz, Bullock-Brennen, Drury-Love},
\begin{equation}
    U^{(4)} = (U^{(2)}_{1}\otimes U^{(2)}_{2}).U^{(4)}_{\mathrm{entangle}}.(U^{(2)}_{3}\otimes U^{(2)}_{4})
    \label{eq:SU(4)-kak}
\end{equation}
where $U^{(2)}_{i}$ are the $SU(2)$ unitary matrices and $U^{(4)}_{\mathrm{entangle}}$ is the $4\times 4$ unitary matrix, which are given by
\begin{align}
& U^{(2)}_{i} = R_{z}(\alpha_{i}) R_{y}(\beta_{i}) R_{z}(\gamma_{i}),\,\,\,\,\,i = 1,..,4\nonumber\\
& U^{(4)}_{\mathrm{entangle}} = e^{i\left( a\,X\otimes X+ b\, Y\otimes Y + c\, Z\otimes Z\right)}
    \label{eq:U-entangle}
\end{align}
with $X,\, Y$ and $Z$ are the Pauli matrices and $a,\, b,\, c$ are the Weyl chamber parameters~\cite{Zhang:2003zz}. The schematic quantum circuit associated with this decomposition is,
\[
    \Qcircuit @C=1em @R=1em{
    \lstick{{q}_{i}: } & \qw & \gate{U^{(2)}_{1}} & \qw & \multigate{1}{U^{(4)}_{\mathrm{entangle}}} & \qw & \gate{U^{(2)}_{3}} & \qw \\
    \lstick{{q}_{i+1}:} & \qw & \gate{U^{(2)}_{2}} & \qw & \ghost{U^{(4)}_{\mathrm{entangle}}} & \qw &\gate{U^{(2)}_{4}} & \qw \\
    }
\]
The entangling gate $U^{(4)}_{\mathrm{entangle}}$ is the following quantum circuit representation,
\[
\Qcircuit @C=1.0em @R=0.7em @!R { 
 \lstick{q_{i}: }& \ctrl{1} & \qw & \ctrl{1} & \gate{H} & \ctrl{1} &\qw & \ctrl{1} &\gate{H} & \gate{\sqrt{\sigma^x}} & \ctrl{1} & \qw & \ctrl{1} & \gate{\sqrt{\sigma^{x}}^\dagger} & \qw \\
 \lstick{q_{i+1}: }& \targ & \gate{R_{z}(-2 c)} & \targ &\gate{H} & \targ &\gate{R_{z}(-2 a)} & \targ &\gate{H} & \gate{\sqrt{\sigma^{x}}}&\targ & \gate{R_{z}(-2 b)} & \targ & \gate{\sqrt{\sigma^x}^\dagger} & \qw   
}
\]
where $R_{z}(\theta)=\mathrm{diag}(e^{-\frac{i \theta}{2}},\,e^{\frac{i\theta}{2}})$. The circuit depth associated with this circuit is 13 which can be minimized by the following optimal circuit\footnote{For previous works see Refs. \cite{Kraus-Cirac, Vatan-Colin, Vidal-Dawson}} \cite{zhang2024optimal} for the $U^{(4)}_{\mathrm{entangle}}$ gate,
\[
\Qcircuit @C=1.0em @R=0.7em @!R { 
 \lstick{q_{i}: }& \targ & \gate{R_z (-2 c )} & \qw & \targ & \gate{R_z ( 2 b)} & \targ & \gate{\sqrt{\sigma^x}} & \qw & \qw & \\
 \lstick{q_{i+1}: }& \ctrl{-1} & \gate{H} & \gate{R_z (-2 a + \frac{\pi}{2})} & \ctrl{-1} & \gate{H} & \ctrl{-1} &  \gate{\sqrt{\sigma^x}^\dagger} & \qw & \qw  
}
\]
whose circuit depth is 7.
\begin{figure}
    \centering
    \includegraphics[width=1.0\textwidth]{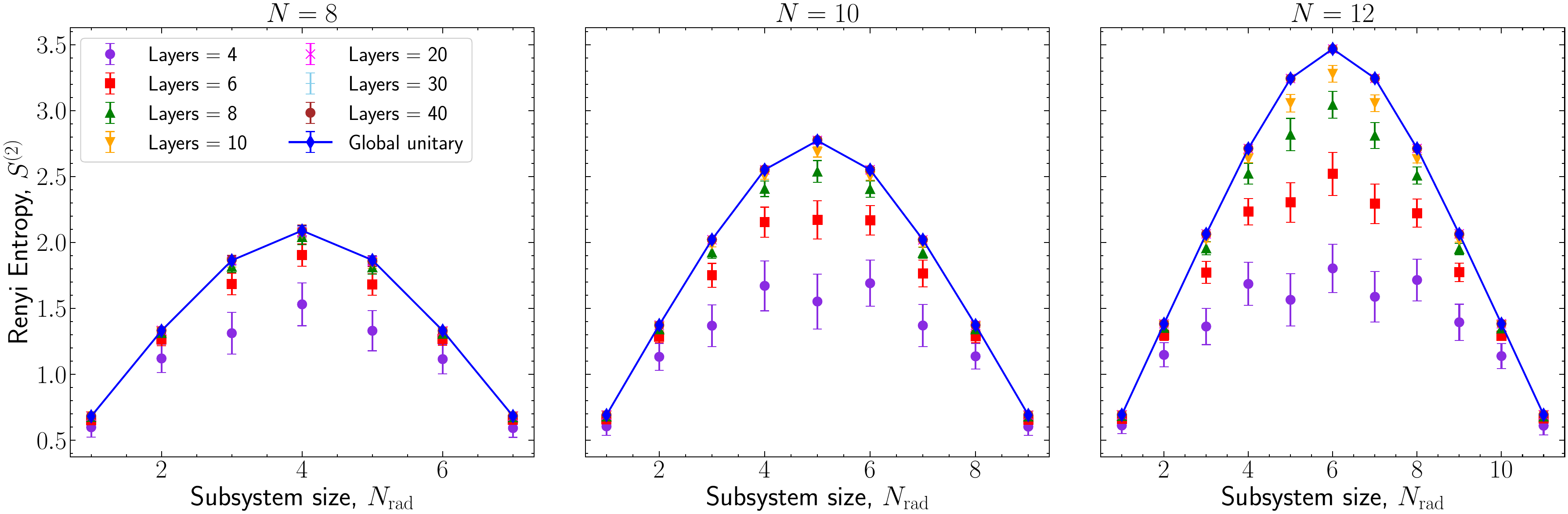}
    \caption{The R\'enyi entropy $S^{(2)}$ of the left-block subsystems of sizes $N_{\mathrm{rad}}$ with increasing number of layers in the brickwall random unitary circuit for three cases $N=8,10,12$. For each case, the average and the standard deviation of the R\'enyi entropy is determined from 100 realizations of the random unitary circuits and the global Haar-random unitary matrices.}
    \label{fig:page-curve-site}
\end{figure}

In Fig. \ref{fig:page-curve-site}, we see the buildup of the entanglement in the subsystems of the $N = 8,\,10$ and 12 qubits with an increasing number of brickwall layers. For all three cases we can see that when the subsystem size is $N_{\mathrm{rad}}=N/2$, i.e., the Hawking radiation qubits to be the half of the black hole qubits, the R\'enyi entropy associated with the subsystem becomes the maximum, whose value depends on the total size of the qubits $N$.

\section{Entanglement entropy measuring protocols in quantum computers}\label{sec:QC-EE}
Consider a quantum computer given access to a quantum state whose purity and R\'enyi entropy we would like to probe. In this regard, we can consider the two protocols to measure the purity and R\'enyi entropy of the corresponding state: 1. swap-based many-body interference protocol and 2. randomized measurement protocol. The following sections present a detailed overview of the two protocols.

\subsection{Swap-based many-body Interference Protocol}\label{sec:swap-protocol}
Consider two systems $A$ and $B$ of $N$ qubits with density matrices $\rho_{A}=|\alpha_{A}\rangle\langle\alpha_{A}|$ and $\rho_{B}=|\beta_{B}\rangle\langle\beta_{B}|$, respectively. Now the action of the swap operation $V_{\mathrm{SWAP}}$ on the states is
\begin{equation}
    V_{\mathrm{SWAP}}|\alpha_{A}\rangle|\beta_{B}\rangle = |\beta_{B}\rangle|\alpha_{A}\rangle .
\end{equation}
Now immediately we can compute the following quantity,
\begin{equation}
    v = \mathrm{Tr}\left[V_{\mathrm{SWAP}}(\rho_{A}\otimes\rho_{B})\right] = \mathrm{Tr}\left[\rho_{A}\rho_{B}\right] \equiv |\langle\alpha_{A}|\beta_{B}\rangle|^{2} .
\end{equation}
We can estimate $v$ using the probability of finding the ancilla qubit in $|0\rangle$ state $P_{0}$, i.e., $v = 2P_{0}-1$, of the following quantum circuit:
\[
\Qcircuit @C=1.0em @R=0.5em @!R{
\lstick{\mathrm{ancilla}: } & \qw & \gate{H} & \ctrl{1} & \gate{H} & \meter\\
\lstick{|\alpha_{A}\rangle: } & \qw & \qw & \qswap & \qw & \qw\\
\lstick{|\beta_{B}\rangle: } & \qw & \qw & \qswap \qwx[-1] & \qw & \qw
}
\]
\vspace{1mm}

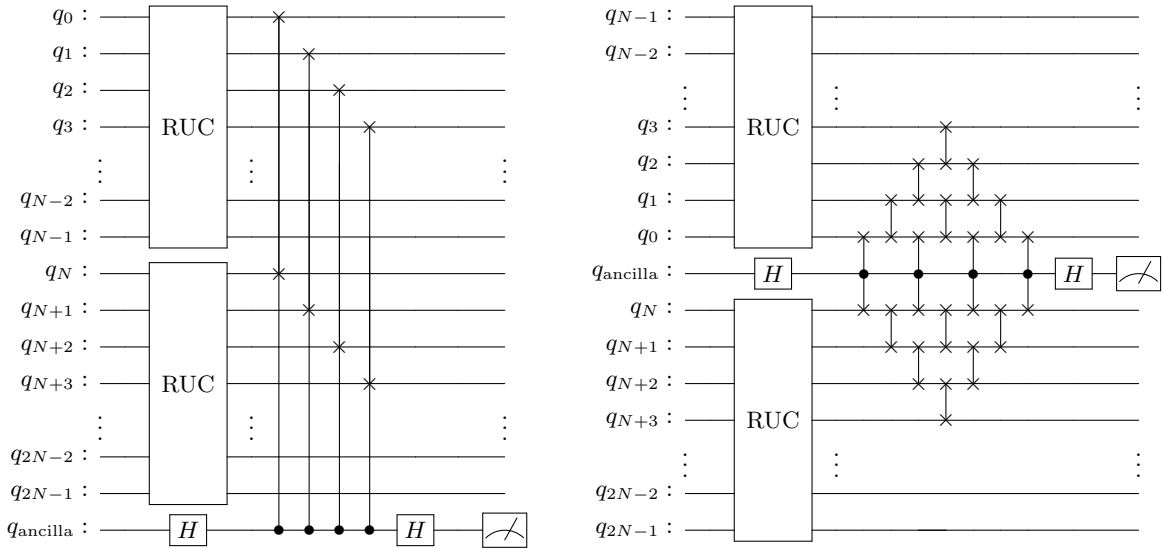
\begin{figure}[t!]
\centerline{
\Qcircuit @C=1.0em @R=0.1em @!R{
\lstick{q_{0}: }   & \qw & \multigate{6}{\mathrm{RUC}} & \qw & \qswap & \qw & \qw & \qw & \qw & \qw &\qw\\
\lstick{q_{1}: }   & \qw & \ghost{\mathrm{RUC}}        & \qw & \qw & \qswap & \qw & \qw & \qw & \qw &\qw\\
\lstick{q_{2}: } & \qw & \ghost{\mathrm{RUC}}        & \qw & \qw & \qw & \qswap  &\qw & \qw & \qw &\qw\\
\lstick{q_{3}: } & \qw & \ghost{\mathrm{RUC}}        & \qw & \qw & \qw & \qw  &\qswap & \qw & \qw &\qw\\
\vdots &  &  & \vdots & & & & & & & \vdots\\
\lstick{q_{N-2}: } & \qw & \ghost{\mathrm{RUC}}        & \qw & \qw & \qw & \qw  &\qw & \qw & \qw &\qw\\
\lstick{q_{N-1}: } & \qw & \ghost{\mathrm{RUC}}        & \qw & \qw & \qw    & \qw & \qw & \qw & \qw &\qw\\
\lstick{q_{N}: }   & \qw & \multigate{6}{\mathrm{RUC}} & \qw & \qswap \qwx[-7] & \qw & \qw & \qw & \qw & \qw &\qw\\
\lstick{q_{N+1}: } & \qw & \ghost{\mathrm{RUC}}        & \qw & \qw    & \qswap \qwx[-6]     & \qw & \qw & \qw & \qw &\qw\\
\lstick{q_{N+2}: } & \qw & \ghost{\mathrm{RUC}}        & \qw & \qw & \qw & \qswap \qwx[-5]  &\qw & \qw & \qw &\qw\\
\lstick{q_{N+3}: } & \qw & \ghost{\mathrm{RUC}}        & \qw & \qw & \qw & \qw  &\qswap \qwx[-4] & \qw & \qw &\qw\\
\vdots &  &  & \vdots & & & & & & & \vdots\\
\lstick{q_{2N-2}: }& \qw & \ghost{\mathrm{RUC}}        & \qw & \qw   & \qw & \qw  &\qw & \qw & \qw &\qw\\
\lstick{q_{2N-1}: }& \qw & \ghost{\mathrm{RUC}}        & \qw & \qw   & \qw & \qw  &\qw & \qw & \qw &\qw\\
\lstick{q_{\mathrm{ancilla}}: }  & \qw & \gate{H}                    & \qw & \ctrl{-14} & \ctrl{-13} & \ctrl{-12} &\ctrl{-11} & \gate{H} &\qw &\meter
}\hspace{2cm}
\Qcircuit @C=1.0em @R=0.1em @!R{
\lstick{q_{N-1}: } & \qw & \multigate{6}{\mathrm{RUC}}        & \qw & \qw & \qw & \qw & \qw & \qw & \qw & \qw & \qw &\qw\\
\lstick{q_{N-2}: } & \qw & \ghost{\mathrm{RUC}}        & \qw & \qw &  \qw & \qw & \qw & \qw & \qw & \qw & \qw &\qw\\
\vdots & & & \vdots & & & & & & & & &\vdots\\
\lstick{q_{3}: } & \qw & \ghost{\mathrm{RUC}}        & \qw & \qw &  \qw & \qw & \qswap & \qw & \qw & \qw & \qw &\qw\\
\lstick{q_{2}: } & \qw & \ghost{\mathrm{RUC}}        & \qw & \qw & \qw  & \qswap & \qswap\qwx[-1] & \qswap & \qw & \qw & \qw &\qw\\
\lstick{q_{1}: }   & \qw & \ghost{\mathrm{RUC}}        & \qw & \qw & \qswap  & \qswap \qwx[-1] & \qswap & \qswap\qwx[-1] & \qswap & \qw & \qw &\qw\\
\lstick{q_{0}: }   & \qw & \ghost{\mathrm{RUC}} & \qw & \qswap &  \qswap \qwx[-1] & \qswap & \qswap\qwx[-1] & \qswap & \qswap\qwx[-1] & \qswap & \qw &\qw\\
\lstick{q_{\mathrm{ancilla}}: } & \qw & \gate{H} & \qw &\ctrl{-1} & \qw & \ctrl{-1} & \qw & \ctrl{-1} & \qw & \ctrl{-1} & \gate{H} & \meter\\
\lstick{q_{N}: }   & \qw & \multigate{6}{\mathrm{RUC}} & \qw &\qswap \qwx[-1] & \qswap & \qswap \qwx[-1] & \qswap & \qswap\qwx[-1] & \qswap & \qswap\qwx[-1] & \qw &\qw\\
\lstick{q_{N+1}: } & \qw & \ghost{\mathrm{RUC}}        & \qw & \qw & \qswap \qwx[-1] & \qswap & \qswap\qwx[-1] & \qswap & \qswap\qwx[-1] & \qw & \qw &\qw\\
\lstick{q_{N+2}: } & \qw & \ghost{\mathrm{RUC}}        & \qw & \qw &  \qw & \qswap \qwx[-1] & \qswap & \qswap\qwx[-1] & \qw & \qw & \qw &\qw\\
\lstick{q_{N+3}: } & \qw & \ghost{\mathrm{RUC}}        & \qw & \qw & \qw & \qw & \qswap\qwx[-1] & \qw & \qw & \qw & \qw &\qw\\
\vdots & & & \vdots & & & & & & & & &\vdots\\
\lstick{q_{2N-2}: }& \qw & \ghost{\mathrm{RUC}}        & \qw & \qw &  \qw & \qw & \qw & \qw & \qw & \qw & \qw &\qw\\
\lstick{q_{2N-1}: }& \qw & \ghost{\mathrm{RUC}}        & \qw & \qw & \qw & \qw & \qw \qw & \qw & \qw & \qw & \qw &\qw
}}
\caption{(Left) Measuring the R\'enyi entropy of the left-subsystem containing $L$ qubits using the SWAP-based many-body interference protocol. Here, using the same random unitary circuit $\mathrm{RUC}$, we prepare two identical random pure states in $\{q_{0},..q_{N-1}\}$ and $\{q_{N},...,q_{2N-1}\}$ qubit registers, respectively. Afterward, we employ the swap operation between qubits $q_{i}\leftrightarrow q_{N+i}$ of two left-subsystems with $L$ (here, $L=4$ as an illustration) qubits and proceed to measurement at the ancilla qubit $q_{\mathrm{ancilla}}$. (Right) To measure the purity and R\'enyi entropy of the left-subsystem of $L$ qubits in a quantum computer with limited qubit connectivity, we rearrange the qubit registers to prepare identical states and use a series of swap gates to bring the qubits $q_{i}$ and $q_{i+N}$ ($i=0,1,.., L$) closest to the $q_{\mathrm{ancilla}}$ before executing the required controlled-swap for the SWAP-MBI protocol.}
\label{fig:swap-protocol}
\end{figure}

Now if one has two identical copies of the $L$-qubit subsystem with reduced density $\rho_{L}$, the swap-based many-body interference protocol presented above will allow us to estimate the purity $\mathrm{Tr}(\rho_{L}^{2})$ and consequently, the R\'enyi entropy, $S^{(2)}=-\mathrm{log}\,\mathrm{Tr}(\rho_{L}^{2})$, as shown in Fig. \ref{fig:swap-protocol}.
In the following steps, we summarize the swap-based many-body interference (SWAP-MBI) protocol to measure the purity and the R\'enyi entropy of a $L$-subsystem within a system of $N$ qubits,
\begin{enumerate}
    \item First, we take a quantum register of $2N+1$ qubits. Then we prepare two identical copies of a quantum state $|\psi\rangle$ by applying the same unitary operator $U$, for example the random unitary circuit $\mathrm{RUC}$ or the time evolution operator $e^{-i H t}$, on the initial state $|i\rangle$ over $\{q_{0},..,q_{N-1}\}$ and $\{q_{N},...,q_{2N-1}\}$ qubit registers, respectively.
    \item Then for our considered subsystem of $L$ qubits $\{q_{i_{1}},...,q_{i_{L}}\}$ where indices $i_{1},...,i_{L}$ are chosen from $[0,N-1]$, we first apply the Hadarmard gate to the ancilla qubit $q_{\mathrm{ancilla}}$ followed by series of controlled-swap on qubits $q_{i_{j}}\leftrightarrow q_{N+i_{j}}$ for $j=1,...,L$, and then another Hadarmard gate to $q_{\mathrm{ancilla}}$. Finally, we measure the ancilla qubit on a computational basis.
    \item If we estimate the probability of finding the $q_{\mathrm{ancilla}}$ in $|0\rangle$ as $P_{0}$ from our measurements, then the purity associated with the $L$-qubit subsystem will be given by $v = \mathrm{Tr}(\rho_{L}^{2}) = 2 P_{0}-1$, and consequently the estimated R\'enyi entropy as $S^{(2)} = -\mathrm{log}(v)$.

    \item For cases of limited qubit connectivity as in Fig. \ref{fig:swap-protocol} (right), when the left-subsystem contains $L$ qubits, the number of swap gates needed to bring $q_{i}$ and $q_{i+N}$ where $i=0,1,..,L$ scales as $L(L-1)$. Besides, the number of controlled-swap gates scales as $L$.
\end{enumerate}

\subsection{Randomized Measurement Protocol}\label{sec:randomized-protocol}
\begin{figure}[t!]
\[
\Qcircuit @C=1.0em @R=0.3em @!R{
\lstick{q_{0}: }   & \qw & \multigate{6}{\mathrm{RUC}} & \qw & \gate{SU(2)} & \meter &\qw\\
\lstick{q_{1}: }   & \qw & \ghost{\mathrm{RUC}}        & \qw & \gate{SU(2)} & \meter & \qw\\
\lstick{q_{2}: } & \qw & \ghost{\mathrm{RUC}}        & \qw & \gate{SU(2)} & \meter & \qw\\
\lstick{q_{3}: } & \qw & \ghost{\mathrm{RUC}}        & \qw & \gate{SU(2)} & \meter & \qw \\
\vdots & & & \vdots & & & \vdots\\
\lstick{q_{N-2}: } & \qw & \ghost{\mathrm{RUC}}        & \qw & \qw & \qw & \qw \\
\lstick{q_{N-1}: } & \qw & \ghost{\mathrm{RUC}}        & \qw & \qw & \qw    & \qw
}
\]
\caption{The randomized measurement (RM) protocol, where the quantum state is prepared using RUC. Then randomized measurements on a subsystem A (here, $L=4$ left-subsystem as an example) are performed by applying $N_{U}$ product of local random unitaries $\hat{U}_{a}=\otimes_{i=1}^{L}U^{(2)}_{i}$ where each $U^{(2)}_{i}$ is sampled from CUE. Subsequently, a measurement in the computational basis is performed. Finally, the statistical correlation of the outcomes of such randomized measurements is used to estimate the purity $\mathrm{Tr}(\rho_{A}^{2})$ and R\'enyi entropy of the reduced density matrix $\rho_{A}$.}
\label{fig:randomized-protocol}
\end{figure}
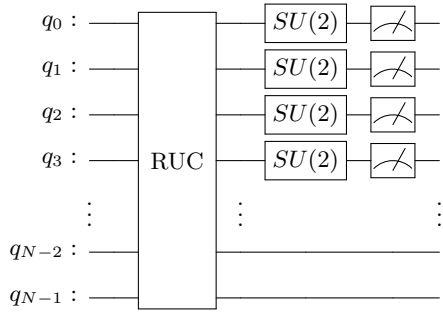

The randomized measurement protocol (RM protocol) uses the fact that the R\'enyi entropy of a quantum system is contained in the statistical correlations between the outcomes of measurement performed on random bases. We are dealing with a pure quantum state $\rho$ of the $N$ qubits produced from the random unitary circuit. Therefore, the purity and the R\'enyi entropy are determined for the reduced density matrices associated with the subsystem $A$ of $L<N$ qubits, in other words, the Hawking radiation qubits. We present a schematic quantum circuit of the RM protocol in Fig. \ref{fig:randomized-protocol}. The experimental implementation of this protocol contains several steps, which we present below. 

First, the quantum state $\rho$ is prepared by applying the random unitary circuit on $N$ qubits. This operation will produce a highly entangled quantum state of the qubits. We aim to measure the purity and R\'enyi entropy associated with the subsystem $A$ of $L$ qubits. Now, one applies a product of single-qubit unitary operators where each of them acts on the $i$-th qubit of the $A$ subsystem, i.e., $i\in A$, as 
\begin{equation}
 \hat{U}_{a}=U^{(2)}_{1}\otimes U^{(2)}_{2}\otimes...\otimes U^{(2)}_{L} .
 \label{eq:single-qubit-local-unitary}
\end{equation}
Additionally, each of the single-qubit unitary $U^{(2)}_{i}$ is drawn independently from the circular unitary ensemble (CUE) of the $SU(2)$. Subsequently, one measures the qubits on the computational basis ($Z$-basis). For each $\hat{U}_{a}$, the repeated measurements are made to obtain the statistics (generally known as shots), which leads one to estimate the occupation probabilities, $P_{\hat{U}_{a}}=\mathrm{Tr}[\hat{U}_{a}\rho\hat{U}^{\dagger}_{a}|j_{L}\rangle\langle j_{L}|]$ of computational basis states $|j_{L}\rangle = |s_{1},s_{2},...,s_{L}\rangle$ with $s_{i}=0,1$. Here, $\hat{U}_{a}$ acts only on the subspace of $L$ qubits. Afterward, the entire process is repeated for $N_{U}$ different randomly drawn instances of $\hat{U}_{a}$.

After determining the set of outcome probabilities $P_{\hat{U}_{a}}(j_{L})$ of the computational basis states $|j_{L}\rangle$ for one instance of random unitaries, $\hat{U}_{a}$, one computes the following quantity,
\begin{equation}
    X_{a} = 2^{L}\sum_{j_{L},j'_{L}}(-2)^{-D[j_{L},j'_{L}]}P_{\hat{U}_{a}}(j_{L})P_{\hat{U}_{a}}(j'_{L})
\end{equation}
where, $j_{L}$ and $j'_{L}$ are different bitstring outcomes for the $L$ qubits, and $P_{\hat{U}_{a}}(j_{L}),\,P_{\hat{U}_{a}}(j'_{L})$ are their corresponding outcome probabilities for the corresponding $\hat{U}_{a}$. In addition, $D[j_{L},j'_{L}]$ is the Hamming distance between bitstrings $j_{L}=s_{1}s_{2}...s_{L}$ and $j'_{L}=s'_{1}s'_{2}...s'_{L}$, measuring how different two bitstrings are, i. e., $D[j_{L},j'_{L}]\equiv \#\{i\in A|s_i\neq s'_i\}$. Consequently the ensemble average of $X_{a}$, denoted by $\overline{X}$,
\begin{equation}
    \overline{X} = \frac{1}{N_{U}}\sum_{a=1}^{N_{U}}X_{a}
    \label{eq:RM-estimated-purity}
\end{equation}
is nothing but the second-order cross-correlations across the ensemble of discrete $N_{U}$ random unitaries $\hat{U}_{a}$, and provides the estimation of the purity $\mathrm{Tr}(\rho_{A}^{2})$ associated with a subsystem of $L$ qubits, $\{q_{i_{1}},...,q_{i_{L}}\}$ where indices $i_{1},..., i_{L}$ are chosen from $[0, N-1]$ in a system of $N$ qubits. Finally, the R\'enyi entropy is 
\begin{equation}
    S^{(2)} = -\mathrm{log}\overline{X}
    \label{eq:renyi-RMP}
\end{equation}

In passing, as shown in \cite{Elben-2}, the RM protocol can be recast as the swap-based many-body interference protocol where the swap operator effectively acts on the two virtual copies of $\rho_{A}$.

\section{R\'enyi entropy of the Hawking Radiation in Quantum Computers}\label{sec:result}

\subsection{Numerical Simulation}\label{sec:numerical-result}
We consider three cases for our numerical simulation: $N=8, 10$, and $12$. First of all, for computing the R\'enyi entropy of a subsystem, we use the simplest, most accurate, but most limited method, which is applying the brickwall layers of SU(4) two-qubit unitary gates in the random unitary circuit Fig. \ref{fig:random-circuit} to the vector of $2^{N}$ complex coefficients, representing the quantum statevector of $N$ qubits. Afterward, we choose the subsystem of $N_{\mathrm{rad}}$ qubits and numerically carry out the partial trace and singular value decomposition to compute the corresponding R\'enyi entropy. The numerical matrix computations corresponding to RUC use QuSpin \cite{quspin} and Qiskit's \texttt{qiskit.quantum$\_$info.Statevector} \cite{qiskit2024}. We denote the results from these computations as `Statevector' in our plots.

To implement the SU(4) two-qubit unitary gates corresponding to the Haar-random $U^{(4)}$, first we determine the $U^{(4)}$ (with $\mathrm{det}U^{(4)}=1$) using the numerical routine \texttt{scipy.stats.unitary$\_$group} prescribed in \cite{Mezzadri}. Then, we use the Cartan-Weyl decomposition Eq. (\ref{eq:U-entangle}) to determine the unitary gates of the two-qubit quantum circuit from $U^{(4)}$. Moreover, we use \texttt{scipy.stats.unitary$\_$group} to determine the global haar-random unitary matrices acting on the $N = 8, 10$, and 12, respectively. In Fig. \ref{fig:page-curve-site}, we can see that the RUC with increasing numbers of brickwall layers can achieve the saturated value of the R\'enyi entropy for different subsystems of $N_{\mathrm{rad}}$ qubits, where the saturated value can be obtained by the respective global Haar-random unitary matrix acting on the $N$ qubits. On the other hand, when $N$ is larger, a larger number of brick wall layers is needed to reach the saturated value of R\'enyi entropy.

For numerical simulation of the two R\'enyi entropy measuring protocols, SWAP-MBI and RM protocols, we use \texttt{qiskit-aer}, a high-performance quantum circuit simulator for Qiskit. For the randomized measurement protocols, the number of unitaries chosen are $N_{U}=10$ and $N_{U}=20$ for $N=8,10$ and $N=12$ cases, respectively. In addition, we take 100 realizations of RUC for each of $N= 8, 10, 12$ cases.

\subsection{IBMQ Results}\label{sec:IBMQ-result}

We implement the RM protocol and SWAP-MBI protocols on the IBM Quantum computer, \texttt{ibm\_marrakesh}, and compare it with our numerical simulation results. \texttt{ibm\_marrakesh} has 156 qubits with basis gates \texttt{CZ, ID, RX, RZ, RZZ, SX, X}. Besides, its lowest two-qubit error rate from all edges measured by isolated randomized benchmarking is about $9.28\times 10^{-4}$, and two-qubit error rate per layered gate (EPLG) for a 100-qubit chain is about $3.71\times 10^{-3}$.

Still, running quantum algorithms on current quantum devices, such as IBM Quantum processors, poses a significant challenge due to the errors and noise present in these systems. To address these issues, quantum error correction (QEC) has been proposed \cite{shor1995scheme, calderbank1996good}. However, implementing QEC comes with a substantial qubit overhead, making it daunting to apply to larger problems, even though they are optimized \cite{kivlichan2020improved, lee2021even}.

Alternatively, quantum error mitigation (QEM) accepts the imperfections of current quantum devices and employs methods to reduce or suppress quantum errors and noise. One of the advantages of QEM is that it often has low or no qubit overhead. Various QEM techniques have been developed in recent years, and their effectiveness has been demonstrated in practical applications \cite{yu2023simulating, Kim-error-mitigation, kim2023evidence, charles2305simulating, chowdhury2024enhancing}.

In our experiments, we apply four QEM methods to mitigate the errors and noise inherent to quantum devices: Zero-Noise Extrapolation (ZNE), Pauli Twirling (PT), Dynamic Decoupling (DD), and Matrix-free Measurement Mitigation (M3). We refer \cite{chowdhury2024enhancing} for further details about these methods. In addition, we adapted an optimized brickwall implementation described in Ref. \cite{zhang2024optimal} which reduces the number of \texttt{CNOT} to $3$ from $6$ in two-qubit unitary gates of RUC. Besides, in the subsequent Qiskit simulations and IBMQ computations, the number of shots is taken as $100000$.

\subsubsection{Randomized Measurement Protocol}\label{sec:randomized-protocol-result}
We measure the R\'enyi entropy using RM protocol for two cases: $N=8$ and $N=12$ for brickwall layers of 4, 6, and 8, respectively. We want to point out that the number of brickwall layers required for RUC to reach the saturation value of the R\'enyi entropy given by the respective global Haar-random unitary matrix for a fixed $N$ is not possible for the current real quantum devices because of their limited maximum coherence time. Therefore, we focus on measuring the buildup of the entanglement with an increasing number of brickwall layers in this work.
\begin{figure}
    \centering
    \includegraphics[width=1.0\textwidth]{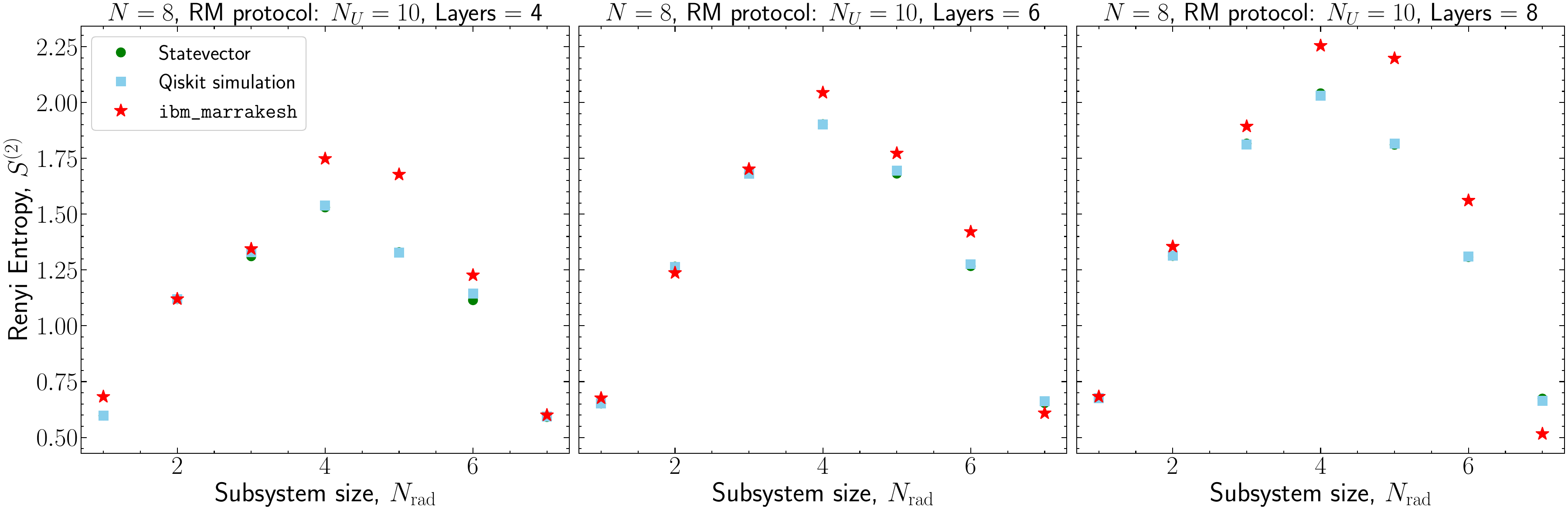}
    \caption{For the system size $N=8$, the R\'enyi entropy $S^{(2)}$ associated with the left-block subsystem, is determined by the randomized measurement protocol (RM protocol) for increasing brickwall layers of the random unitary circuit. Here, the number of local unitaries is set to $N_{U}=10$. Besides, for both Statevector computation and Qiskit simulation of the RM protocol, we use 100 realizations of the random unitary circuit, respectively.}
    \label{fig:page-curve-N8-randomized}
\end{figure}

\begin{figure}
    \centering
    \includegraphics[width=1.0\textwidth]{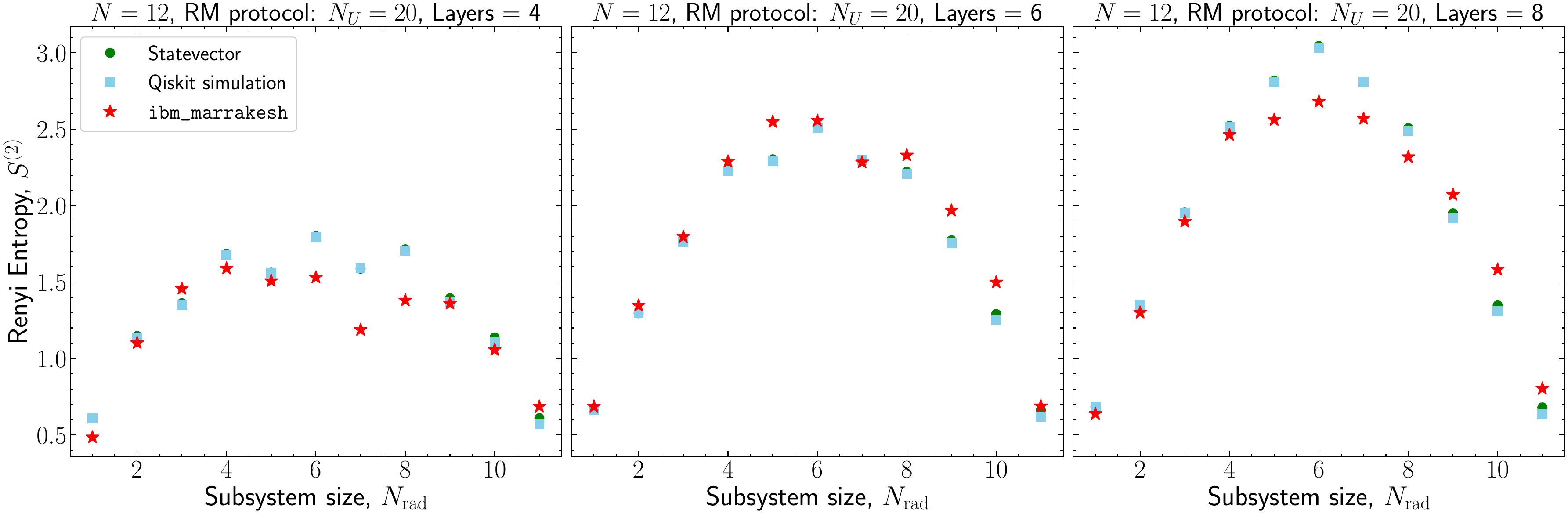}
    \caption{For the system size $N=12$, the R\'enyi entropy $S^{(2)}$ associated with the left-block subsystem, is determined by the randomized measurement protocol (RM protocol) for increasing brickwall layers of the random unitary circuit. Here, the number of local unitaries is set to $N_{U}=20$. Besides, for both Statevector computation and Qiskit simulation of the RM protocol, we use 100 realizations of the random unitary circuit, respectively.}
    \label{fig:page-curve-N12-randomized}
\end{figure}

In Fig. \ref{fig:page-curve-N8-randomized} and \ref{fig:page-curve-N12-randomized}, we see that there is a good agreement between the R\'enyi entropy for different subsystem sizes $N_{\mathrm{rad}}$, obtained with the statevector (exact) computation, and the RM protocol using Qiskit simulation (noiseless) and \texttt{ibm\_marrakesh}. Here, we choose the number of unitaries $N_{U}=10$ and $N_{U}=20$ for $N=8$ and 12, respectively. 

The corresponding circuit depths associated with RUC for brickwall layers 4, 6, and 8 are 126, 181, and 235 for both $N=8$ and 12 cases. However, the \texttt{CNOT} gate count turns out to be $84, 126, 168$ for brickwall layers 4, 6, and 8, respectively with $N=8$ whereas for $N=12$ it is $132, 198, 264$, respectively.  For ten brickwall layers or more, due to the limited coherence time of the actual device, the results, even with error mitigation, deviate significantly from the theoretical values. Therefore, in real device measurement, we limit ourselves to a maximum of 8 brickwall layers in RUC.

\subsubsection{Swap-based many-body interference protocol}\label{sec:swap-protocol-result}
We also implement the swap-based many-body interference protocol, for the first time, in \texttt{ibm\_marrakesh}, a superconducting quantum computer. To overcome the limited qubit connectivity of the IBM device, we use the series of swap gates to bring the corresponding qubits closest to the ancilla qubit and execute the controlled-swap gate as described in Fig. \ref{fig:swap-protocol} (right).
\begin{figure}
    \centering
    \includegraphics[width=1.0\textwidth]{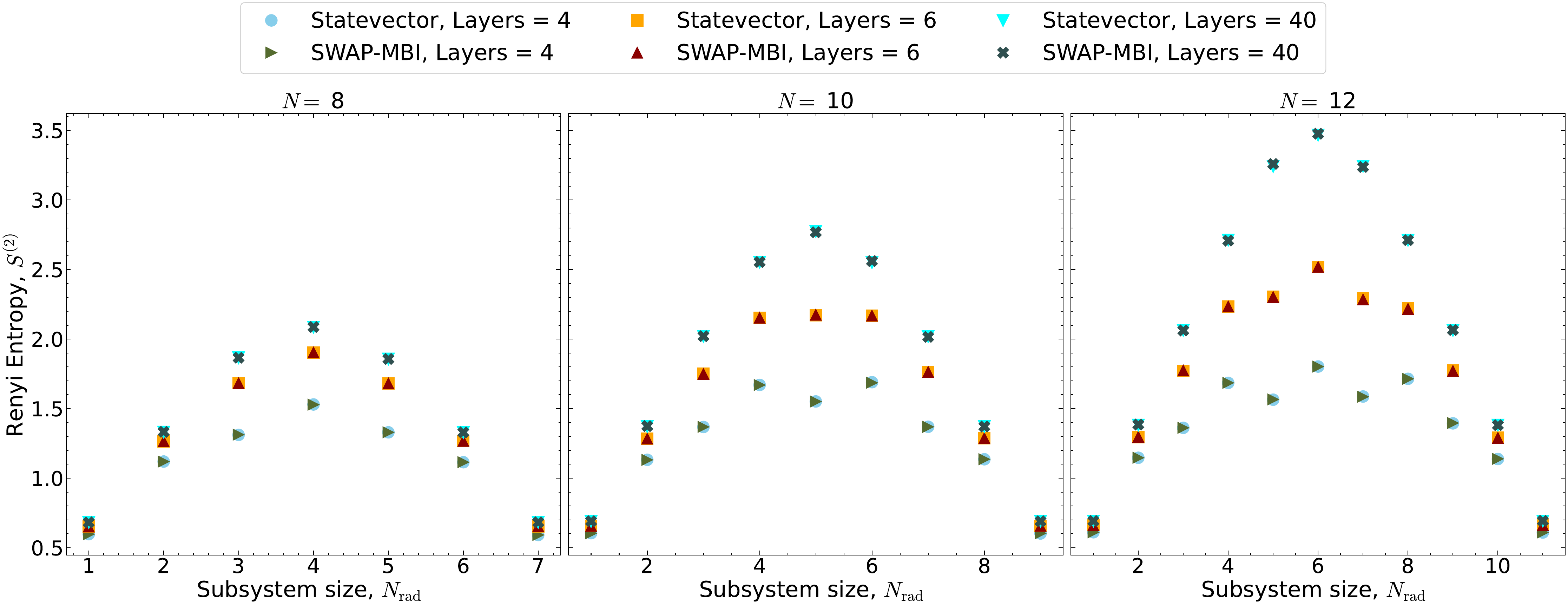}
    \caption{The R\'enyi entropy $S^{(2)}$ with respect to the subsystem size $N_{\mathrm{rad}}$ for different systems $N = 8, 10$ and 12, determined by the swap-based many-body interference (SWAP-MBI) protocol. Here, for both Statevector computation and Qiskit simulation of the SWAP-MBI protocol, we use 100 realizations of the random unitary circuit, respectively.}
    \label{fig:page-curve-swap-protocol}
\end{figure}

\begin{figure}[t!]
    \centerline{\includegraphics[width=0.33\textwidth]{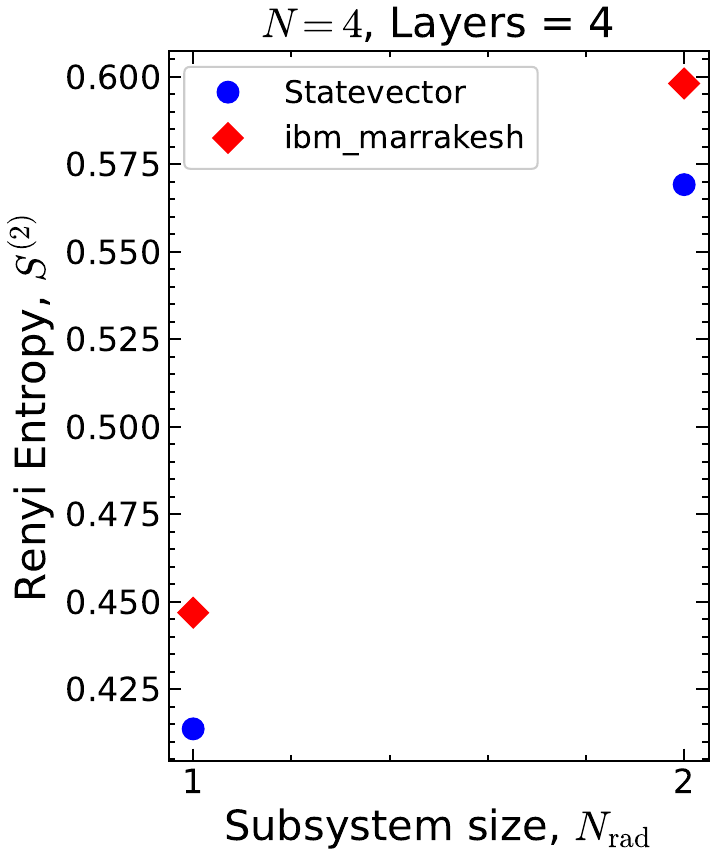}\hspace{1mm}\includegraphics[width=0.33\textwidth]{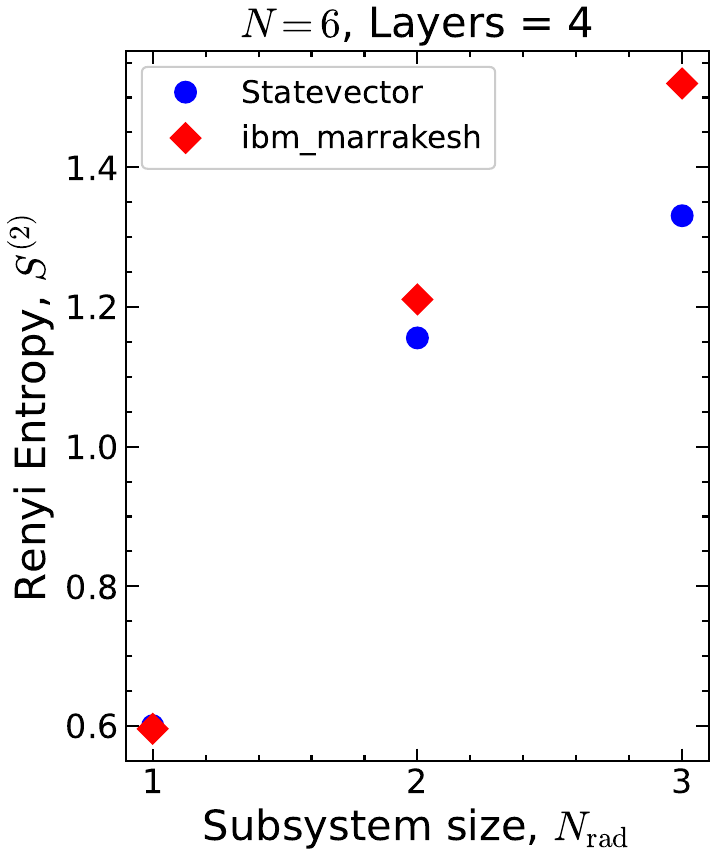}\hspace{1mm}\includegraphics[width=0.33\textwidth]{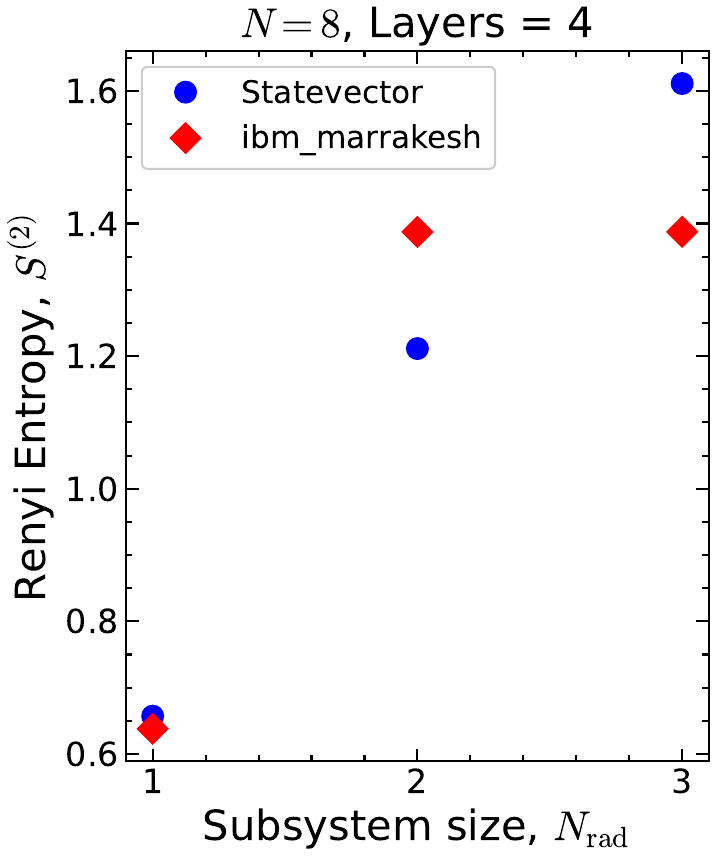}}
    \caption{The R\'enyi entropy $S^{(2)}$ with respect to the subsystem size $N_{\mathrm{rad}}$ for different systems $N = 4, 6$ and 8 at four brickwall layers, determined by Statevector computation and swap-based many-body interference (SWAP-MBI) protocol at \texttt{ibm\_marrakesh}. Here, we only consider single but different realizations of RUC for each $N$.}
    \label{fig:swap-device}
\end{figure}
As we see from Fig. \ref{fig:page-curve-swap-protocol}, the numerical results from the Qiskit simulation of SWAP-MBI protocol and Statevector (exact) computations match very well. However, the results from \texttt{ibm\_marrakesh} disagree with the Statevector results when $N_{\mathrm{rad}}$ increases. We see from Fig. \ref{fig:swap-device} that the R\'enyi entropy of $N_{\mathrm{rad}}=1$ subsystem agrees with the numerical values for $N=4$, 6 and 8, but the results differ significantly from theoretical values for $N_{\mathrm{rad}}=3$. Actually, the implementation of the SWAP-MBI protocol leads to more circuit depths and \texttt{CNOT} counts compared to the RM protocol. In all cases of $N=4,\,6,\, 8$ in Fig. \ref{fig:swap-device}, the circuit depths are roughly $\{176, 231, 281\}$ for the subsystem size $N_{\mathrm{rad}}=1,2,3$, respectively. The quantum circuit executing the SWAP-MBI protocol's swap test involves circuit depths of $\{55,110,160\}$ for the subsystem sizes of the $N_{\mathrm{rad}}=1,2,3$, respectively. Therefore, we can see a significant increase in the circuit depth for the SWAP-MBI protocol compared to the RM protocol, which causes the SWAP-MBI protocol to perform poorly in the IBM devices. Moreover, the noise from the quantum computers could make the two copies of the quantum state deviate, contributing to the error. For this reason, the SWAP-MBI protocol warrants developing more optimized quantum circuits and investigations, which we set for future works. 

\section{Conclusion and Outlook}\label{sec:conclusion}
The investigation of black hole information dynamics through the perspective of quantum computing can provide a significant avenue for addressing one of the crucial questions in modern theoretical physics: the resolution of the black hole information paradox. However, establishing the usefulness of this rapidly evolving technology for such investigation requires rigorous testing and validation of a simplified toy model of black hole evaporation, namely the qubit-transport model. In this respect, our work focused on whether the current noisy quantum computer, such as IBM's superconducting quantum computer, can capture the entanglement dynamics in such a toy qubit model of black hole evaporation. Firstly, we simulate the black hole's scrambling dynamics by implementing an efficient random unitary circuit. Afterward, by incorporating quantum error mitigation methods in entropy measurement protocols, we obtain the R\'enyi entropy associated with the qubit subsystems within current hardware limitations, which are in excellent agreement with the numerical results. Thus, we demonstrate that the current noisy quantum computers can effectively model the evolution of entanglement entropy within the toy qubit model, such as the qubit transport model of black hole evaporation.

In conclusion, we illustrate that current quantum computers are on the right path to tackle significant challenges in physics. Our work opens avenues for further studies into more sophisticated models of black hole evaporation and the eventual development of large-scale, error-corrected quantum simulations of black hole physics. We have seen that near-term quantum computers can simulate the scrambling dynamics of a black hole by implementing random unitary operations. However, in the future, fault-tolerant quantum computers will be capable of going even further by simulating and verifying proposals such as the non-isometric holographic model of black holes~\cite{Akers:2022qdl, Kim:2022pfp}, a model that connects the interior of a black hole to its exterior and is suggested as a potential resolution to the black hole information paradox. This research showcases the potential of quantum computers to elucidate the complexities of black hole information dynamics and bring us closer to a comprehensive understanding of quantum gravity in the future.

\section*{Acknowledgments}
T.A.C would like to thank the High Energy Theory group in the Department of Physics and Astronomy at the University of Kansas for their hospitality and support where this work has been done. R.S.S. is supported by Laboratory Directed Research and Development (LDRD No. 23-051) of Brookhaven National Laboratory and RIKEN-BNL
Research Center. This work is supported by the U.S. Department of Energy, Office of Science, Grants No. DE-SC0012704 (K.Y.) and the Brookhaven National Laboratory LDRD No. 24-061 (K.Y.).
This research used quantum computing resources of the Oak Ridge Leadership Computing Facility, which is a DOE Office of Science User Facility supported under Contract DE-AC05-00OR22725. 
This research used resources of the National Energy Research Scientific Computing Center, a DOE Office of Science User Facility supported by the Office of Science of the U.S. Department of Energy under Contract No. DE-AC02-05CH11231 using NERSC award DDR-ERCAP0028999.

\end{document}